\documentclass[prl, amsfonts, twocolumn, nofootinbib, showpacs, letter]{revtex4}
\usepackage{graphicx, epsfig}

\begin{document}
\title{Birkhoff's theorem fails to save MOND from non-local physics}
\author{De-Chang Dai$^{1}$, Reijiro Matsuo$^{2}$, Glenn Starkman$^{2}$}
\affiliation{$^{1}$Department of Physics, SUNY at Buffalo, Buffalo, NY 14260-1500}
\affiliation{$^{2}$CERCA, Department of Physics, Case Western
Reserve University, Cleveland, OH~~44106-7079}


\begin{abstract}

 \widetext
We investigate the consequences of Birkhoff's theorem in general relativity (GR) and in Modified Newtonian dynamics (MOND). 
We study, in particular, the system of a finite-mass test particle inside a spherical shell. 
In both GR and MOND, we find non-vanishing acceleration for that test particle.  
The direction of the acceleration is such that it pushes the test particle toward the center of the shell. 
In GR, the acceleration is found analytically in the case of a small test mass with a small displacement
from the center of the shell. 
In MOND, the acceleration is found analytically in the limit of large test mass and small displacement,
and a comparison to numerical values is made. 
Numerical simulations are done for more general cases with parameters that mimic the system of a galaxy in a cluster.
In GR, the acceleration is highly suppressed, and physically insignificant. 
In MOND, on the contrary, the acceleration of the point particle can be a significant fraction of
the field just outside of the spherical shell.
\end{abstract}


\maketitle

\section{\bf Introduction}
\indent

Birkhoff's theorem (BT) states that in General Relativity (GR) any  spherically symmetric solution
of the Einstein field equations in vacuum must be stationary and  asymptotically flat.  
As a consequence the  metric exterior to a spherically symmetric mass distribution
must be a Schwarzschild metric.  One important corollary of the theorem is that the metric inside a spherical shell
(or inside the innermost of a sequence of concentric such sheels) is the Minkowski metric. 

Birkhoff's Theorem is the generalization from Newtonian theory to GR of Gauss' Law for gravity.
As a consequence of Gauss'  Law, outside a spherically symmetric mass distribution,
the Newtonian gravitational field is that of a point mass at the center of the distribution. 
Meanwhile, the gravitational field anywhere inside any spherical mass shell vanishes, 
In the context of the inverse-square-law of Newtonian gravity, these results are
easily understood. For example, at a point inside a spherical  shell, the force on a
test particle from any thin ring on the shell, is precisely balanced by the force
due to a ring subtending the same angle but directly  opposite the first ring.
Even if the test particle is not at the center. 
This is because the decrease in the gravitational force due to one ring being farther away from the test particle,
is precisely balanced by the increase in area (and so mass) of that farther-away ring.

One can think of Birkhoff's theorem as describing the motion of a zero-mass test particle
in the presence of a spherically symmetric mass distribution. 
A zero-mass test particle does not spoil the spherical symmetry.
However, in realistic situations, the ``test particles'' probing a gravitational field have a finite mass.
Often, this mass is not small at all, as in the case of a large galaxy inside a cluster.
Therefore, unless the test particle is at the center of the distribution,
its presence perturbs the system from spherical symmetry, spoiling the assumptions underlying BT.
This paper addresses the impact this violation of spherical symmetry has both in GR
and in theories that implement the ideas of Modified Newtonian Dynamics (MOND).

MOND \cite{MOND1, MOND2, Milgrom1}, the reader will recall, is an alternative to dark matter in which the missing gravity
problem of galaixes is solved by altering the gravitational force law, rather than by
introducing new unseen forms of matter.

In both GR and MOND, there are reasons to suspect that the force on a test particle inside a spherical mass shell
may not actually vanish.
General relativity is a metric theory. 
Massive bodies change the geometry around them.
A massive test particle distorts the space around it
and thus a "spherical shell" ceases to be a spherical shell unless the test particle is at its center.
This distortion effect from test particle will appear in any modification of GR that preserves its metric nature  --
such as DGP, TeVeS \cite{Bekenstein1}, Generalized Einstein Aether (GEA) and f(R). 

In the case of MOND,  which is not itself a metric theory (although its covariant implementations, such as TeVeS and  GEA are),
the inverse-square law which is crucial for the vanishing acceleration of the test particle inside the shell
(or at least for our intuitive understanding of why it vanishes) does not exist. 
MOND was proposed to explain galactic rotation curves,
and therefore is characterized by a gravitational field scaling as $r^{-1}$ instead of $r^{-2}$
at large distances from a bounded mass distribution. 

\begin{figure}
    \centering{
    \includegraphics[width=2.5in]{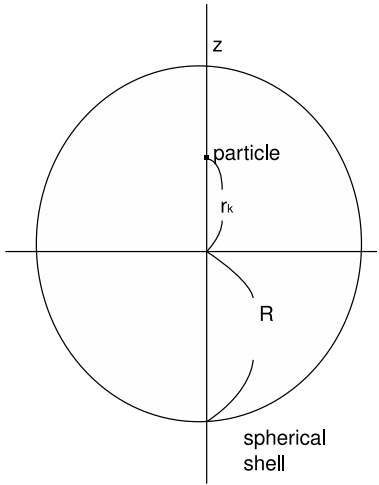} }
    \caption{Mass distribution: We consider a point particle inside a large spherical shell.  R denotes the radius of the shell, and $r_{k}$ denotes the amount of displacement of the particle from the center of the shell.}
\label{dist}
\end{figure}

In what follows, we consider the mass distribution shown in Fig. \ref{dist} --
a spherical shell of mass $M_s$ plus an interior off-center point mass, $M_{k}$, of much lower mass.
We may consider this distribution to be a very simplified version of for example a galaxy in a cluster,
or a clump of matter -- star, globular cluster, clump of dark matter -- inside a galaxy.
For definiteness we will the total mass of the system to be typical of a galaxy cluster, $M_{total}=2.0\times10^{14}M_{\odot}$,
and take the radius of the shell to be fixed at a typical cluster raius, $R=2.0 Mpc$. 
We now proceed to investigate the acceleration of the point mass in both General Relativity and MOND.


\section{General relativity}

In Newtonian mechanics, a spherical shell is a well-defined object.
However, in GR, since particles change the geometry around them,
the simple definition of a sphere as a set of point equidistant from some common point (the center) is lost
in the presence  of a perturbing mass.
We define a spherical shell using Parametrized Post-Newtonian (PPN) coordinates.
We identify a point, the center, such that the distance from the center in PPN coordinates
is the same for each point on the shell.    We call this distance $R$ the  radius of the shell.

We can calculate the acceleration of the test particle in GR from the Einstein-Infeld-Hoffman (EIH) equation \cite{EIH}:
\begin{eqnarray}
\frac{d^2 x_{k}}{dt^{2}}&=&\frac{dv_{k}}{dt}=\sum_{a\neq k}\vec{r}_{ak}\frac{GM_{a}}{r_{ak}^{3}}\biggl[1-4\sum_{b\neq k}\frac{GM_{b}}{r_{bk}}\nonumber\\
&-&\sum_{c\neq a}\frac{GM_{c}}{r_{ca}}(1-\frac{\vec{r}_{ak}\cdot \vec{r}_{ca}}{2r_{ca}^{2}})+v_{k}^{2}+2v_{a}^{2}\nonumber \\
&-&4\vec{v_{a}}\cdot \vec{v_k}-\frac{3}{2}\biggl(\frac{\vec{v}_a \cdot \vec{r}_{ak}}{r_{ak}}\biggr) ^2 \biggr] \\
&-&\sum_{a\neq k}(\vec{v}_{a}-\vec{v}_{k})\frac{GM_{a}\vec{r}_{ak}\cdot (3\vec{v}_{a}-4\vec{v}_{k})}{r_{ak}^{3}}\nonumber\\
&+&\frac{7}{2}\sum_{a\neq k}\sum_{c\neq a}\vec{r}_{ca}\frac{G^{2}M_a M_c}{r_{ak}r_{ca}^{3}} \quad . \nonumber \label{Einstein}
\end{eqnarray}
Here $\vec{r}_{a}=\vec{x}_{a}-\vec{x}$ and  $\vec{r}_{ab}=\vec{x}_a-\vec{x}_b$.
$\vec{x}$ is the spatial variable in the PPN coordinate frame, and $\vec{x}_{a}$ is location of object $a$.
This equation describes the free-fall of each point in the system.
In order to describe the acceleration of the test particle in our system,
we should remove from equation (\ref{Einstein}) the terms that are associated with the acceleration of the shell
(which is much smaller than the acceleration of the point particle).
The acceleration is  then
\begin{eqnarray}
\frac{d^2 x_{k}}{dt^{2}}&=&\frac{dv_{k}}{dt}\simeq-\sum_{a\neq k}\vec{r}_{ak}\frac{G^{2}M_{a}M_{k}}{r_{ak}^{4}} \label{gracceleration} \\
&=&-\frac{G^{2}M_{k}M_{s}}{4r_{k}^{2}R}\biggl[\ln \left(\frac{R-r_{k}}{R+r_{k}}\right)+\frac{2r_{k}R}{(R^2-r_{k}^2)}\biggr]\hat{z} \quad .\nonumber
\end{eqnarray}
In obtaining the final result, we have replaced the sum by an integral over the shell which we were able to perform.
Notice that there is a singularity as $r_k\rightarrow R$.
This singularity is not real.
It appears because PPN is not valid as $r_k$ is very close to the shell.

Defining $d=r_{k}/R$, $R_{Schs}=2GM_{s}$ and $g_{Ns}=\frac{GM_{s}}{R^{2}}$,
equation (\ref{gracceleration}) can be written as
\begin{eqnarray}
\frac{d^2 x_{k}}{dt^{2}}&=&-g_{Ns}\frac{M_{k}}{M_{s}}\frac{R_{Schs}}{R}\frac{1}{4d}\left[\frac{1}{1-d^{2}}+\frac{1}{2d}\ln\left(\frac{1-d}{1+d}\right)\right]\hat{z} \nonumber \\
                        &\simeq&-g_{Ns}\frac{M_{k}}{M_{s}}\frac{R_{Schs}}{2R}\left(\frac{1}{3}d+\frac{2}{5}d^{3}+\frac{3}{7}d^{5}+\cdots\right)\hat{z} ,
\end{eqnarray}
where the last line is an expansion for small $d$.

The gravitational acceleration clearly does not vanish -- it points toward the center.
A particle inside a spherical shell is therefore attracted toward the center to restore spherical symmetry, 
where the acceleration happily does vanish.
However, since $M_{k}\ll M_{s}$,  the acceleration is very small compared to, say, the Newtonian acceleration
just outside  the shell, $g_{Ns}$.
It is suppressed by both the ratio of the  test mass to the shell mass, and, more importantly,
by the ration of the Schwarzschild radius $R_{Schs}$ of the shell to its physical radius. 
This suppression factor is huge for typcial astrophysical systems,
$9.7\times10^{-6}$ for our choice of mass and radius. 
This suppression factor makes the acceleration physically insignificant in GR.  

\section{MOND}
In 1984 Bekenstein and Milgrom introduced the Lagrangian formulation of Modified Newtonian dynamics (MOND) \cite{Bekenstein}. 
The field equation of MOND is derived from the Lagrangian
\begin{equation}
L=-\int d^{3}r\left\{\rho\psi+(8\pi G)^{-1}a_{0}^{2}\mathcal{F}\left[\frac{(\nabla\psi)^{2}}{a_{0}^{2}}\right]\right\} \quad , \label{Lagrangian}
\end{equation}
where $\psi$ is the gravitational potential.
$\mathcal{F}(y^{2})$, with $y\equiv|\nabla\psi|/a_{0}$, is an arbitrary universal function,
that together with $a_{0}$, the characteristic scale of MOND,  specifies the theory.
Varing L with respect to $\psi$ yields a modified Possion equation:
\begin{equation}
\nabla\cdot[\mu(|\nabla\psi|/a_{0})\nabla\psi]=4\pi G\rho(x) \quad , \label{ModPossion}
\end{equation}
where $\mu(y)\equiv\mathcal{F}'(y^{2})$. 
$\mu(y)$ must approach 1 as $|y|\gg 1$ and $|y|$ as $|y|\ll 1$,
in order that the field scale as $\frac{1}{r^{2}}$ near a spherical  mass distribution (the usual Newtonian result)
and as $\frac{1}{r}$ far from the mass distribution to explain flat galaxy rotation curves. 

One the widely used form of $\mu$ is
\begin{equation}
\mu(y)=\frac{|y|}{(\sqrt{1+|y|^{2}}}) \label{mufunction}\quad .
\end{equation}
(However see \cite{Famaey} for different form of $\mu$ function.) 
The value of $a_{0}$ is then given by phenomenological fit. 
We will adapt the value derived by Begeman et al \cite{Begeman}
in the study of external galaxies with high quality rotation curves.
\begin{equation}
a_{0}=1.2\times10^{-10}m/s^{2}\quad . \nonumber
\end{equation}

Again we consider the mass distribution in Fig. \ref{dist}. 
The center of the spherical shell is chosen to be the center of coordinates. 
Because of the axial symmetry of the configuration, we put the point particle on the z-axis.

We assume that the radius of the shell is large enough that the gravitational acceleration goes deep into the MOND regime. 
For a bounded mass distribution of total mass $M$, we define a transition radius
\begin{equation}
R_{t}=\sqrt{GM/a_{0}} \quad .
\end{equation}
$R_{t}$ indicates a point at which the Newtonian field approximately equals $a_{0}$,
and this is about the point at which the field switches from Newtonian $1/r^2$ to MOND's $1/r$. 
Thus we take $R\gg R_{t}$.
For the physical parameters we are considering, $R/R_{t}=4.2$.

Non-linearity in the modified Possion equation negates the superposition principle,
and it makes an analytic calculation of the field a non-trivial task. 
We present the perturbative solution for the case $M_{k}\gg M_{s}$,
and in more general cases, we resort to numerical calculations. 

\section{Perturbative calculation}

We first consider the mass of the particle inside the shell to be much larger than the mass of the shell. 
In this case, the density distribution of the spherical shell can be treated as an aspherical perturbation
on a spherical system -- the point mass --
and a perturbative solution to the modified Possion equation is possible.
Let the density distribution of the point particle and the spherical shell be $\rho_{k}$ and $\rho_{shell}$, respectively. 
Then the unperturbed field $\psi_{0}$ satisfies the modified Poisson equation:
\begin{equation}
\nabla\cdot[\mu(|\nabla\psi_{0}|/a_{0})\nabla\psi_{0}]=4\pi G\rho_{k} \quad .\nonumber
\end{equation}
The exact solution for $\nabla\psi_{0}$ can be found by applying Gauss's theorem. 
In terms of the quantity $u=\frac{GM_{k}}{a_{0}|\vec{r}-\vec{r_{z}}|^{2}}$,
\begin{eqnarray}
\left|\frac{\nabla\psi_{0}}{a_{0}}\right|&=&\sqrt{\frac{1}{2}u^{2}+\sqrt{\frac{1}{4}u^{4}+u^{2}}} \\
                                   &\simeq&u^{\frac{1}{2}}(1+\frac{1}{4}u+...) \quad .                   \nonumber
\end{eqnarray}
The expansion in small $u$ is valid for $|\vec{r}-\vec{r_{k}}|\gg\sqrt{\frac{GM_{k}}{a_{0}}}$. 
Knowing $\nabla\psi_{0}$, one can find the first-order perturbation equation for $\psi_{1}$ using
\begin{equation}
\nabla\cdot[|\nabla\psi_{0}|/a_{0}((\nabla\psi_{1}\cdot\hat{e_{0}})\hat{e_{0}}+\nabla\psi_{1})]=4\pi G\rho_{shell}\quad,  \nonumber
\end{equation}
where $\hat{e_{0}}$ is a unit vector pointing in the direction of $\nabla\psi_{0}$.  In the deep MOND regime where $\mu(|\nabla\psi_{0}|/a_{0})\sim |\nabla\psi_{0}|/a_{0}$, and in the limit $R\gg r_{k}$, one can expand $\psi_{1}$ just inside and outside of the shell in powers of $\frac{r_{k}}{R}$.  The first order aspherical contribution to the potential is
\begin{equation}
\psi_{1}=\left\{ \begin{array}{c}\frac{r_{k}}{4}\sqrt{\frac{Ga_{0}}{M_{k}}}M_{s}\frac{\cos\theta}{r} \mbox{  for $r\leq R$}\\  
               \frac{r_{k}}{4}\sqrt{\frac{Ga_{0}}{M_{k}}}M_{s}\frac{r\cos\theta}{R^{2}} \mbox{  for $r>R$} \end{array} \right. \quad .
\end{equation}
The net force on the spherical shell due to $\psi_{0}$ and $\psi_{1}$ is
\begin{eqnarray}
\vec{F}_{shell}&=&\left\{\frac{M_{s}\sqrt{GM_k}a_{0}}{2r_{k}}\biggl[ 1+\frac{R^{2}-r_{k}^{2}}{2r_{k}R}\ln\biggl(\frac{R-r_{k}}{R+r_{k}}\biggr)\biggr] \right.  \nonumber \\
         &-&\frac{M_{s}}{8}\frac{(GM_{k})^{\frac{3}{2}}}{a_{0}^{\frac{1}{2}}}\frac{1}{r_{k}}\left[\frac{1}{R^{2}-r_{k}^{2}}+\frac{1}{2r_{k}R}\ln\left(\frac{(R-r_{k})}{(R+r_{k})}\right)\right] \nonumber \\
         &-&\left.\frac{5}{24}\frac{r_{k}}{R^{2}}\sqrt{\frac{Ga_{0}}{M_{k}}}M_s^{2} + \cdots\right\} \hat{z} \quad . \label{Sforce}
\end{eqnarray}
The net force on the shell points toward positive $\hat z$,
having the  effect of restoring the spherical symmetry by moving the shell so that the
it is centered on the point mass.
The first two terms result from $\psi_{0}$ and the last term comes from $\psi_{1}$. 
We note that the first term, which is dominant over other terms,
is independent of the specific choice of $\mu$ function. 
It is a consequence solely of the MOND condition: $\mu(y)\rightarrow y$ for $y\ll1$. 

The Lagrangian (\ref{Lagrangian}) from which the field equation (\ref{ModPossion}) is derived
is invariant under spacetime translations and spatial rotations. 
Energy-momentum and angular momentum conservation therefore hold for an isolated system in MOND. 
In particular, Newton's third law  of action and reaction is valid. 
Hence, the acceleration $a_k$ of the point particle in the frame of the shell
can be found from equation (\ref{Sforce}). 
Writing $R_{k}=\sqrt{GM_{k}/a_{0}}$, $d=r_{k}/R$, and expanding in a Taylor series,
\begin{eqnarray}
\frac{a_{k}}{a_{0}}&\simeq&\left\{-\frac{M_{s}}{M_{k}}\left(\frac{R_{k}}{R}\right)\left(\frac{1}{3}d+\frac{1}{15}d^{3}+\frac{1}{35}d^{5}+...\right)\right. \nonumber \\
                   &+&\frac{M_{s}}{M_{k}}\left(\frac{R_{k}}{R}\right)^{3}\left(\frac{1}{12}d+\frac{1}{10}d^{3}+\frac{3}{28}d^{5}+...\right) \nonumber \\
                   &+&\left.\frac{5}{24}\left(\frac{M_{s}}{M_{k}}\right)^{2}\left(\frac{R_{k}}{R}\right)d\right\}\hat{z} \quad .
\end{eqnarray} 
As expected, $a_{k}$ approaches zero as $d\rightarrow0$.
For $d\neq 0$, $a_{k}$ points toward the center of the spherical shell.
As the point particle is displaced, the system reacts to restore the spherical symmetry. 
We see here that when the system as a whole is not spherically symmetric,
the acceleration on a test particle in a spherical shell is non-vanishing.

The suppression of the acceleration compared to characteristic accelerations in the system
is much milder in MOND than that in GR. 
In MOND, the scale of the acceleration is $a_{0}$, but  suppressed by $R_{t}/R$,
whereas in general relativity, it the scale is $g_{N}$, but suppressed by $R_{Sch}/R$. 
Since $R_{t}\gg R_{Sch}$ in general and since $a_{0}\gg g_{N}$ for a system deep in the MOND regime,
the scale acceleration is much larger in MOND than in GR. 
For $M_{total}=2.0\times10^{14}M_{\odot}$ and $R_{Sch}=2GM_{total}$,
the ratio of accelerations is approximately $(a_{0}\times R{t})/(g_{N}\times R_{Sch})=4.3\times 10^{5}$.

\section{Numerical calculation}
To find the acceleration of the point particle in a more general case,
we adapt the numerical scheme developed by Milgrom \cite{Milgrom}. 
In this scheme, the field $U$ is defined by
\begin{equation}
\vec{U}=\mu(|\nabla\psi|/a_{0})\nabla\psi \label{ufunction}\quad .
\end{equation}
So long as $\mu$ is monotonic,
one can invert the relation in equation (\ref{ufunction}) and write $\nabla\psi$ in terms of U.
For  the specific choice of $\mu$ given by equation (\ref{mufunction}),
\begin{equation}
\nabla\psi=\sqrt{\frac{1}{2}+\sqrt{\frac{1}{4}+\frac{a_{0}^{2}}{|U|^{2}}}}\vec{U} \quad .
\end{equation}
The field $\vec{U}$, then, satisfies the set of differential equations
\begin{eqnarray}
\vec{\nabla}\cdot \vec{U}=4\pi G\rho       \nonumber \\
\vec{\nabla}\times\sqrt{\frac{1}{2}+\sqrt{\frac{1}{4}+\frac{a_{0}^{2}}{|U|^{2}}}}\vec{U}=0 \quad .
\end{eqnarray}
The discretization and numerical calculation on the lattice is done in terms of $\vec{U}$.
Because of the one-to-one correspondence between $\nabla\psi$ and $\vec{U}$, one can find $\nabla\psi$ from $\vec{U}$.

An initial ansatz for $\vec{U_{i}}$ in the numerical solution is given by solving
\begin{equation}
\nabla\cdot U_{i}=4\pi G\rho \quad .\nonumber
\end{equation}
The field $\vec{U_{i}}$ has the correct divergence, but not the correct curl. 
The code iterates to make the curl vanish at each vertex of the lattice.
$\vec{U_{i}}$ also serves as the boundary condition for the numerical solution. 
Since $\nabla\cdot(\vec{U}-\vec{U_{i})}=0$,
$\vec{U}$ and $\vec{U_{i}}$ differ only by the curl field. 
It can be shown \cite{Bekenstein} that the curl field for a bounded mass distribution
vanishes at least as fast as $\sim \frac{1}{r^{3}}$. 
Then, assuming that the physical size of the lattice is large compared to the mass distribution,
$U\rightarrow U_{i}$ on the boundary.

We implement Milgrom's algorithm on a spherical lattice.
The point particle is placed at the center of the lattice,
and the center of the spherical shell is displaced from the center of the lattice by $-r_{k}$. 
The total number of angular grid lines is denoted by L,
and the total number radial grids is fixed at $L/5$. 
To meet the boundary condition, the radius of the outermost shell is set to be $100R$.

In the case $M_{k}\gg M_{s}$, a comparison between the numerical results and the perturbative solution is possible. 
We make the comparison between the solutions with fixed values of $M_{s}/M_{k}=0.01$ and $R_{k}/R=0.25$.
\begin{figure}
\centering{
\includegraphics[width=3.0in]{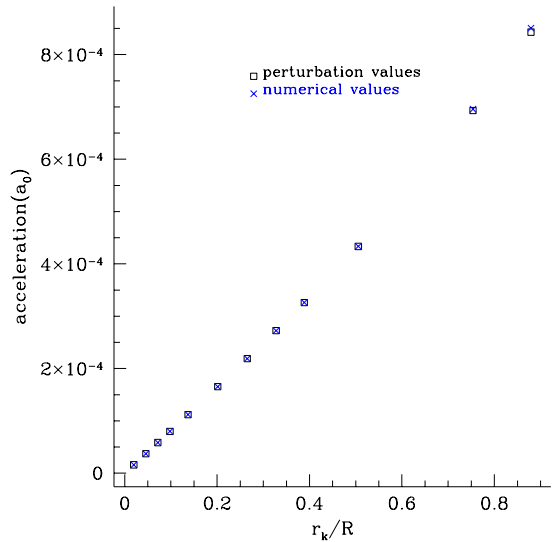}}
\caption{This figure plots the acceleration on the point particle as a function of $d=r_{k}/R$.  They all points toward the center of the spherical shell.  The numerical solutions and the corresponding perturbation solution is shown.}
\label{comp1}
\end{figure}
\begin{figure}
\centering{
\includegraphics[width=3.0in]{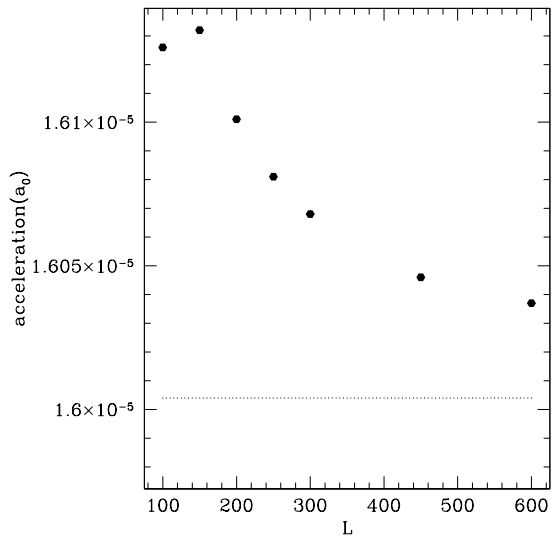}}
\caption{This figure shows the dependence of the numerical solution on the number of lattice sites.  The value of d is fixed at $d=0.02$.  The dotted line indicates the value from the perturbation solution.}
\label{comp2}
\end{figure}
Fig. \ref{comp1} shows that these agree to better than one percent. 
On the plot shown, the percentage difference varies from $0.1\%$ to $0.9\%$.
The discrepancy between the values increases as $d=r_{k}/R$ approaches 1,
where the perturbative expansion becomes less reliable. 
Fig. \ref{comp2} shows how numerical values depend on the number of lattice points. 
As expected, the values approach the perturbative solution as L increases. 
In the following study, we use the lattice size of $L=300$ and $L=600$.  

\begin{figure}
\centering{
\includegraphics[width=3.5in]{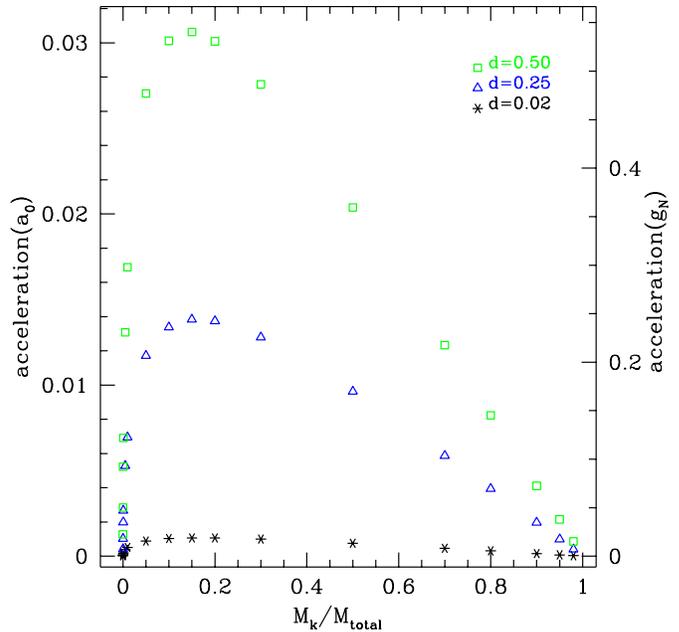}}
\caption{This figure shows the acceleration of the point particle as a function of $M_{k}/M_{total}$
for particles at a variety of positions. 
The acceleration points toward the center of the shell. 
The numerical results are for L=300.}
\label{mass}
\end{figure}
\begin{figure}
\centering{
\includegraphics[width=3.5in]{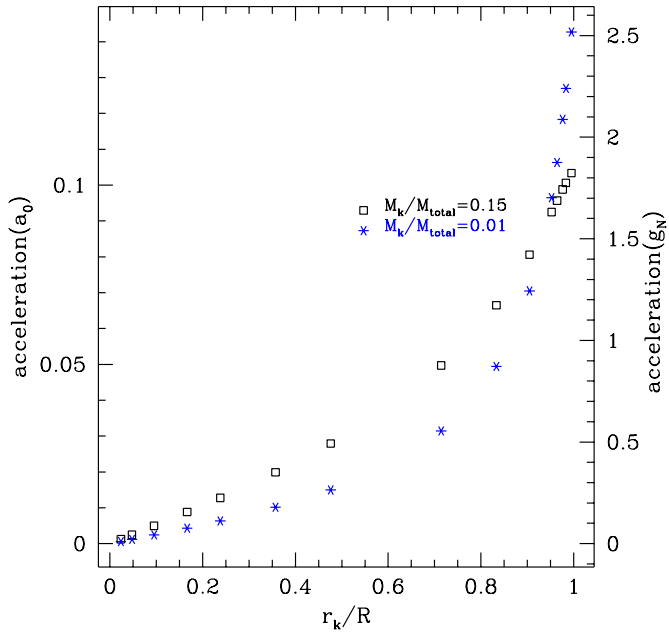}}
\caption{This figure shows the acceleration on the point particle as a function of position for fixed particle-mass to total-mass ratio. Results are plotted for two different mass ratio. The acceleration points toward the center of the shell.  Simulations are performed on the lattice with L=600.}
\label{aposition}
\end{figure}

Fig. \ref{mass} and Fig. \ref{aposition} represent the results of the numerical calculations. 
The mass dependence of the acceleration of the point particle is plotted in Fig. \ref{mass},
and the position dependence is plotted in Fig> \ref{aposition}. 
We see clearly that there is non-vanishing acceleration directed toward the center of the spherical shell. 
In Fig. \ref{mass}, we see that the acceleration vanishes in both the $M_{k}\rightarrow 0$ limit and as $M_{s}\rightarrow 0$.
This is expected since these are two limits in which spherical symmetry is recovered. 
For the three curves plotted in Fig. \ref{mass},
the peak value occur around $M_{k}/M_{total}\sim0.15$. 
The position of the peak has a slight position dependence --
as the particle gets closer to the shell (large d),
the peak occurs at smaller $M_{k}/M_{total}$ ratio. 

We note one curious feature from Fig. \ref{mass}. 
The astronomically interesting region in Fig. \ref{mass} is where $M_{k}/M_{s}\ll1$. 
In this regime, the acceleration of the point particle is a very sharp function of its mass. 
This implies that galaxies with slight mass differences might experience quite different acceleration in theoriest that
implement the MOND limit.

The distance dependence of the acceleration is shown in Fig. \ref{aposition}. 
Again, the acceleration of the particle vanishes when the particle is near the center ($d\rightarrow 0$),
and increases monotonically outward.
Near the shell ($d\sim1$),
the value for $M_{k}/M_{total}=0.01$ exceed that for $M_{k}/M_{total}=0.15$. 
This is because the peak mentioned above occurs at lower value of $M_{k}/M_{total}$ for larger value of $d$. 

Unlike for GR, in MOND the acceleration of the point particle is a significant fraction of $g_{N}$. 
Especially when the particle is close to the shell, the acceleration can be larger than $g_{N}$.     

\section{Concluding remarks and implications}
In both general relativity and MOND,
the acceleration of a massive test particle inside a spherical mass shell vanishes only when
the particle is at the center of the shell. 
When the particle is displaced from the center,
the particle experiences a force toward the center of the shell,
the direction in which spherical symmetry would be restored. 
The magnitude of the acceleraion in GR is not physically significant in most (or probably all)
astrophysical situations of interest,
since it is suppressed by the ration of the Schwarzschild radius of the shell to the size of the shell.

In MOND, on the contrary, the acceleration of the point particle is a significant fraction of the
surface gravity just outside the shell.
This is despite the fact that MOND has a Birkhoff or Gauss-like theorem which implies that the
potential inside an isolated spherical mass shell is constant.  
However,  this  theorem operates in the absence of the usual explanation for Gauss' Law in
Newtonian gravity or classical electrostatics -- the balance between the $r^2$ growth of surface
areas and the $r^{-2}$ force law.  Thus, when the very particular conditions of Birkhoff's
theorem are broken even a little, the graviational force re-emerges at considerable strength.
The characteristic MOND field for the values of parameters used in this paper is approximately
$\frac{\sqrt{GM_{total}a_{0}}}{R}=0.24a_{0}$.  From Fig. \ref{mass} and Fig. \ref{aposition},
we can see that the acceleration inside the spherical shell can be significant at this scale. 

In MOND, the applicability of Birkhoff's theorem is very limited, and in computations
one needs to consider not only the local mass distribution but also the background mass distribution.  

\section{acknowledgements}
We want to thank Irit Maor for helpful suggestion in development of the code.  We also wish to thank Yi-Zen Chu for helpful discussion regarding EIH equation.
GDS and RM are supported by a grant from the Department of Energy in support of the Particle Astrophysics theory group. DCD is supported by grants from the HEPCOS group at SUNY.

\end{document}